\documentstyle[epsf]{mn}

\newcommand{\etal}{{\rm et al.~}}
\newcommand{\Mpc}{$h^{-1}$~{\rm Mpc}}

\title[Oscillating correlation function]
      {The supercluster--void network II.\\
      An oscillating cluster correlation function}

\author[J. Einasto et al.]
  { J. Einasto$^1$,  M. Einasto$^1$, P. Frisch $^2$, S.
  Gottl\"ober$^3$, V. M\"uller$^3$, V. Saar$^1$, 
  \newauthor A.A. Starobinsky$^4$, 
  E. Tago$^1$,  D. Tucker$^{3,5}$, and H. Andernach$^{6,7}$ \\
  $^1$ Tartu Observatory, EE-2444 T\~oravere, Estonia \\
  $^2$ G\"ottingen University Observatory, D-37083 G\"ottingen, Germany \\ 
  $^3$ Astrophysical Institute Potsdam, An der Sternwarte 16,
       D-14482 Potsdam, Germany \\
  $^4$ Landau Institute Theoretical Physics, Moscow, Russia \\
  $^5$ Fermilab, Batavia, IL 60510, USA \\
  $^{6}$ INSA; ESA IUE Observatory, E--28080 Madrid, Spain \\
  $^7$ Depto. de Astronom\'\i a, Univ. Guanajuato, Guanajuato, 
       Mexico }

\date{Accepted  Received  ; in original form 1996}
\pubyear{1997}

\begin{document}

\maketitle

\begin{abstract} 

We use rich clusters of galaxies in the Northern and
Southern Galactic hemispheres up to a redshift $z=0.12$ to determine
the cluster correlation function for a separation interval $\approx 650$
\Mpc\ ($h$ is the Hubble constant in units of 100~${\rm km~s^{-1}~Mpc^{-1}}$).
We show that superclusters of galaxies and voids between them form a
moderately regular network. As a result the correlation function determined
for clusters located in rich superclusters oscillates: it has a series
of regularly spaced secondary maxima and minima. The scale of the
supercluster-void network, determined from the period of oscillations,
is $P=115\pm 15$ \Mpc. Five periods are observed.  The correlation
function found for clusters in poor and medium rich superclusters is
zero on large scales.  The correlation functions calculated separately
for the Northern and Southern Galactic hemispheres are 
similar. The amplitude of oscillations for clusters in the Southern
hemisphere is larger by a factor of about 1.5.
 
We investigate the influence of possible errors in the correlation
function. The amplitude of oscillations for clusters in very rich
superclusters is about 3 times larger than the estimated error. 
We argue that the oscillations in the correlation function are neither
due to the double-cone shape of the observed volume of space, nor to
the inaccuracy in the selection function.

We compare the observed cluster correlation function with similar
functions derived for popular models of structure formation, as well
as for simple geometrical models of cluster distribution. We find 
that the production of the observed cluster correlation function in
any model with a smooth transition of the power spectrum from a
Harrison-Zeldovich regime with positive spectral index on long
wavelengths to a negative spectral index on short wavelengths is
highly unlikely. The power spectrum must have an extra peak located at
the wavelength equal to the period of oscillations of the correlation
function.  The relative amplitude of the peak over the smooth spectrum
is probably of the order of a factor of at least 1.25.  

These quantitative tests show that high-density regions in the Universe
marked by rich clusters of galaxies are distributed more regularly
than expected. Thus our present understanding of structure formation
needs revision.

\end{abstract}


\section{Introduction}

A fundamental property of the distribution of galaxies is clustering,
manifested by the presence of groups and clusters of galaxies and
quantitatively measured by the correlation function. Due to clustering the
correlation function of galaxies has a large positive value at small
separations. At a separation of $\sim 30$ \Mpc\ the correlation function
approaches (or crosses) zero and remains small on larger scales. A correlation
function of zero has been interpreted as an indication of a random
distribution of galaxies.  This picture -- clustering on small scales and a
random scale-free distribution on larger scales -- formed the classical
paradigm of the large-scale distribution of galaxies and clusters of galaxies.

The discovery of superclusters consisting of clusters and filaments of
galaxies and huge voids between them has changed this classical
paradigm.  According to available data superclusters reside in chains
and walls, separated by voids of diameters of about 100 \Mpc, and form
a rather regular network (Einasto \etal 1994, Einasto \etal 1997c,
hereafter Paper I).  This raises a question about the existence of
some regularity in the distribution of superclusters of galaxies, and,
if so, about the presence of a related scale in the Universe.

The first clear demonstration for the possible presence of a
regularity in the distribution of galaxies on very large scales came
from a deep pencil-beam survey of galaxies by Broadhurst \etal (1990).
This survey covers small areas near the North and South Galactic poles
and has a depth of about 700 \Mpc\ in both directions.  The galaxy
density shows periodic peaks separated by $\sim$128 \Mpc.  In total
over 10 peaks have been observed. Bahcall (1991) explained
high-density regions in the distribution of galaxies by the presence
of superclusters.

There has been much discussion regarding the implication of this
result.  Kaiser and Peacock (1991) have argued that a peak in the
one-dimensional spectrum can arise without any large-scale feature in
the three-dimensional distribution of galaxies. Dekel \etal (1992)
investigated the problem and showed that this periodicity is barely
compatible with Gaussian fluctuations in the framework of CDM-type
scenarios of structure formation. Thus, the initial reaction to the
observation of Broadhurst \etal was that there is no need to change
the classical paradigm on the distribution of matter on large scales.

However, other independent data on the possible presence of some
regularity in the distribution of matter on large scales in the
Universe have accumulated.  In the 70ies Shvarzman and Kopylov
initiated a program to study the large-scale distribution of matter.
They used Abell (1958) clusters of galaxies of richness $R\geq 2$, and
rich, compact clusters from the list of Zwicky \etal (1961--69);
redshifts were determined for clusters up to $z\approx 0.3$ in a
region around the Northern Galactic pole.  This survey indicated the
presence of a secondary peak in the correlation function at $\approx
125$ \Mpc\ (Kopylov \etal 1984, 1988). Later the survey was extended
to the Southern Galactic hemisphere, and a peak in the correlation
function on the same scale was found (Fetisova \etal 1993).  Mo \etal
(1992a, b) and Einasto \& Gramann (1993) used a different method to
analyse the cluster correlation function, and the presence of a
feature at $\sim 130$ \Mpc\ was confirmed.  Similar scale was found in
the distribution of clusters using other methods like the void and
pencil-beam analysis (Einasto \etal 1994, Paper I).

Landy \etal (1996) derived the 2D power spectrum of the Las Campanas
Redshift Survey and found a peak at a wavelength 100 \Mpc. The peak is
due to numerous density enhancements located at this characteristic
mutual separation.  The same redshift survey was analysed also by
Tucker \etal (1995, 1997) and Doroshkevich \etal (1996) who also found
characteristic features on similar scales.  A $\sim
100$ \Mpc\ scale has also been seen in the distribution of QSO absorption
line systems (Quashnock, Vanden Berk \& York, 1996).

During the past few years the number of redshifts determined for rich
clusters of galaxies has rapidly increased. This makes a new analysis
of cluster data worthwhile, as the Abell-ACO cluster sample is the
deepest almost full-sky survey available at present. In this paper we
study the correlation function for clusters of galaxies using a recent
compilation of available data on clusters of galaxies by Andernach,
Tago, \& Stengler-Larrea (1995, 1997).  Our study follows approaches by
Bahcall and Soneira (1983) and more recently by Peacock and West
(1992) and Einasto \etal (1993). However, in contrast to all previous
studies we concentrate here on large scales, i.e. well beyond 100~\Mpc. To do
this we consider the whole dataset of clusters now available for both
the northern and southern Galactic hemispheres as a single sample of
depth $\approx 700$~\Mpc.  The same dataset has been used in Paper I
to derive a  catalogue of superclusters of galaxies and to study
the spatial distribution of clusters, by Jaaniste \etal (1997) to
investigate the orientation and shape of superclusters of galaxies,
and by Saar \etal (1995) to determine the correlation function with a
novel method.  Methodical problems connected with the determination
and interpretation of the correlation function on large scales are
discussed separately by Einasto \etal (1997b, hereafter Paper
III). The power spectrum for our cluster sample was found and
discussed by Einasto \etal (1997a, hereafter E97).

The  paper is structured as follows.  In  Section 2 we
describe the observational data used and the selection functions of
the data. Section 3 is devoted to the analysis of the correlation
function of clusters of galaxies on large scales. We determine the
correlation function for the whole sample as well as for subsamples of
clusters in the Northern and Southern Galactic hemispheres, and for
cluster populations located in rich and poor superclusters.  In
Section 4 we discuss the influence of the smoothing length, inaccuracy
of the selection function, and other factors on our results.  In
Section 5 we compare our results with simulations using simple
geometrical models and results of $N$-body calculations for the CDM
model and a double power-law model. In Section 6 we derive the
possible cluster power spectrum from models.
A summary of the main results is given in Section 7.

We use a Hubble constant of $H_0 = 100~h {\rm km~s^{-1}~Mpc^{-1}}$.

\section{DATA} \label{data}

The Abell--ACO catalogue of clusters of galaxies (Abell 1958, Abell,
Corwin, \& Olowin 1989) is presently the largest available source of
the large-scale distribution of matter in the Universe covering the
whole sky outside the Milky Way zone of avoidance. We use for the
present study a recent compilation of measured redshifts for these
clusters by Andernach, Tago, \& Stengler-Larrea (1997). This
compilation gives redshifts for a total of about 2000 Abell--ACO
clusters (including supplementary, or S-clusters). We used the 1995
version of the compilation, omitted all S-clusters and used only
clusters with measured redshifts up to $z=0.12$.  To this sample we
added all clusters with photometric redshift estimates $z_{est}\le
0.12$.  Our full sample contains 1304 Abell--ACO clusters of galaxies,
869 of which have measured redshifts.  

\begin{figure*}
\vskip 2cm
\epsfysize=7cm 
\epsfbox [-10 260 290 500]{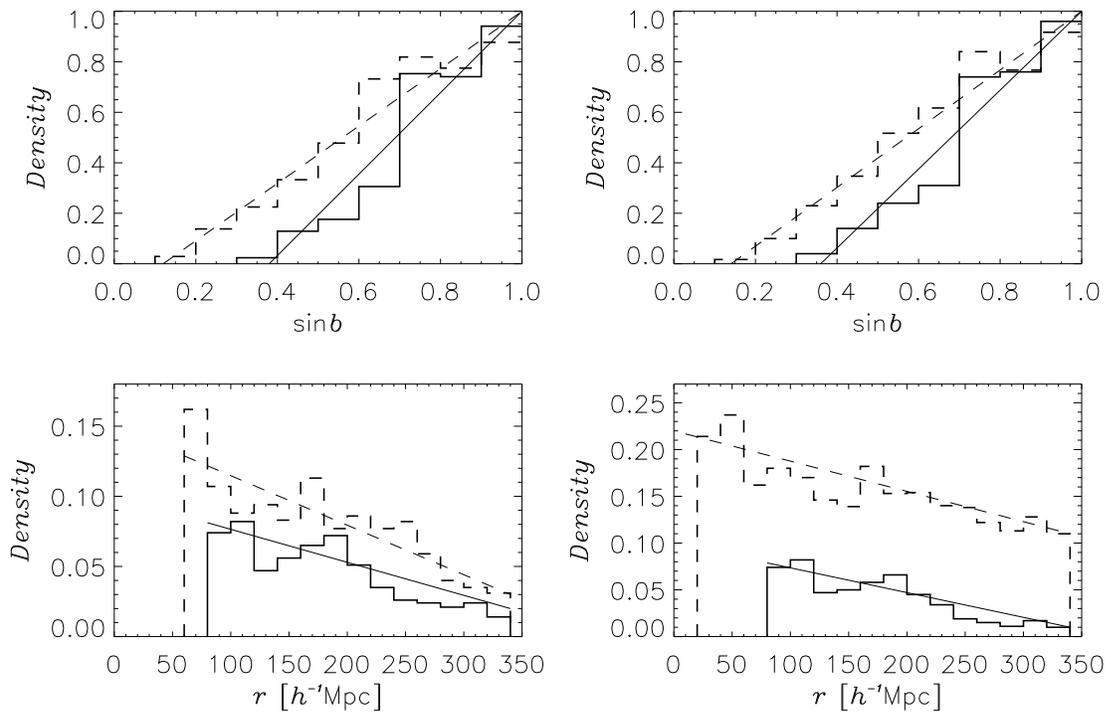}
\caption{
  Selection functions for clusters of galaxies.  The volume density of
  clusters is shown as function of the sine of Galactic latitude $b$ (upper
  panels) and as function of distance $r$ from the observer (lower panels). In
  the upper panels the density is given in units of the density near the
  Galactic pole (i.e. $\sin b = 1$); in the lower panels in arbitrary units. In
  the left panels only clusters with measured redshifts were used; in the
  right panels we used all clusters.  Dashed lines are for clusters located in
  low-density environments (isolated clusters and clusters in superclusters
  with less than 8 members); solid lines are for high-density regions
  (clusters in superclusters with at least 8 members). Dashed and solid
  straight lines represent linear approximations of the selection function.}
\label{fig1}
\end{figure*}

We have included clusters of richness class 0 in our study. About half
of all clusters in the nearby region studied are of this richness
class and the number of objects is crucial in the present work.  Abell
clusters of richness class 0 are X-ray emitters and hosts of cD
galaxies with extended haloes as often as clusters of higher richness.
Both facts suggest that these clusters are physical objects which can
be used to trace the large-scale structure.  Possible projection
effects discussed by Sutherland (1988), Dekel \etal (1989) and others
are not crucial for the present study as we are mostly interested in
the distribution of clusters on large scales.  A small excess of
cluster pairs on small separations noted by Sutherland and Dekel \etal
can be considered as an additional selection effect. 

This sample was used in Paper I to derive a new catalogue of
superclusters and to study their spatial distribution.  In the
present paper we use both the cluster sample and the supercluster
catalogue.  The use of the supercluster catalogue gives us the
possibility to analyse the distribution of clusters in different
environments. Superclusters were determined using a
``friends-of-friends'' technique with neighbourhood radius
24~\Mpc. This radius was chosen on the basis of the multiplicity
function which shows that individual superclusters start to become
evident at a neighbourhood radius of about 16
\Mpc; at radii larger than 30 \Mpc\ superclusters begin to join into
huge agglomerates with dimensions exceeding the characteristic scale
of the supercluster-void network. Thus the neighbourhood radius must
lie within these boundaries. The influence of this radius on our
results for the correlation function shall be studied below (Section
4.3).

\begin{table}
\begin{minipage}{120mm}
\caption{ Selection function parameters}
\halign to \hsize {# \hfil &\quad\hfil #&\quad \hfil#\hfil&\quad\hfil
#\hfil&\quad \hfil#\hfil&\quad \hfil#\hfil\cr
\noalign {\smallskip}
Sample&$s_0$&$d_{0N}$&$d_{1N}$&$d_{0S}$&$d_{1S}$\cr
\noalign {\medskip}
ACO.R.H8    & 0.38&1.00&0.80&1.00&0.80\cr
ACO.R.L8    & 0.12&1.00&0.80&1.00&0.80\cr
ACO.A.H8    & 0.36&1.00&0.50&1.00&0.50\cr
ACO.A.L8    & 0.14&0.78&0.36&1.00&0.52\cr
}
\end{minipage}
\end{table}

In Paper I superclusters were divided into richness classes according
to their multiplicity (the number of member clusters in
superclusters).  It was also shown that the overall distribution of
superclusters of different richness is rather similar: superclusters
are located in chains that form a fairly regular network. The mean
diameter of voids between superclusters is $\sim 100$ \Mpc.  The
skeleton of the supercluster-void network is formed by very rich
superclusters. Poor and medium rich superclusters as well as isolated
clusters are scattered around them, leaving void interiors empty of
rich clusters. The distribution of superclusters in void walls depends
on the supercluster richness: the mean separation between poor and
medium rich superclusters is small and has a smooth distribution
whereas the separation between very rich superclusters is much larger
and its distribution is peaked: over 75~\% of very rich superclusters
are located at separations $110 - 150$ \Mpc\ on opposite sides of
voids.

This finding motivated us to study the correlation function of clusters of
galaxies located in superclusters of different richness.  As in Paper I we
divide cluster samples into populations using the supercluster richness as the
parameter which determines {\it the mean density of the large-scale
  environment} of clusters (see Frisch \etal 1995).  In contrast to Paper I we
divide superclusters into only two richness classes with variable richness
threshold. We shall use the following nomenclature of cluster samples. The
first 3 capital letters ACO denote clusters from the Abell--ACO catalogue
(excluding S-clusters); the following capital letter indicates whether we use
the sample of all clusters (A) or the sample of clusters with measured
redshifts (R); the following capital letter denotes cluster samples in high-,
or low-density environments (respectively H or L); the last number indicates
the limiting multiplicity $N_{cl}$ of superclusters used to divide the sample
into high- and low-density populations.  Clusters belonging to superclusters
with {\it at least} $N_{cl}$ members were attributed to the high-density
population, and isolated clusters as well as clusters in superclusters with
{\it less} than $N_{cl}$ members to the low-density population.

To calculate the correlation function of clusters of galaxies we 
generate Poisson samples of test particles with the same shape and
selection function as the real samples.  The selection effects depend
on Galactic absorption, on the difficulty to find lower richness 
clusters at large distances, on the decrease in the fraction of
clusters with measured redshifts with distance, the differences in the
mean density of clusters in the Abell and ACO catalogues, etc.

Poisson samples must be generated with all these effects taken into account.
We have calculated the selection function as a function of two variables, the
Galactic latitude $b$, and the distance from the observer $r$, separately for
the Northern and Southern Galactic hemispheres. We determined selection
functions for clusters populating rich and poor superclusters, using a
threshold richness of $N_{cl}=8$.  The influence of the choice of the
threshold richness $N_{cl}$ shall be discussed in the next Section.

In Figure~1 we show the results of the determination of the selection function
for clusters of galaxies with measured redshifts.  The number of clusters vs.
the Galactic latitude was determined as a function of $\sin b$. Differences
between the two hemispheres are small, thus in Figure~1 we present the mean of
both hemispheres.  Data are normalised to unit density at $\sin b =1$. We see
an almost linear decrease of the number density of clusters with $\sin b$.
This linear regression, $D(b)=(\sin b - \sin b_0)/(1-\sin b_0)$, is given by
the value $s_0=\sin b_0$ where the density of cluster reaches 0, and it was
used to calculate Poisson samples for the correlation function.

\begin{figure*}
\epsfysize=7cm
\vskip -17mm
\epsfbox [-10 30 300 260]{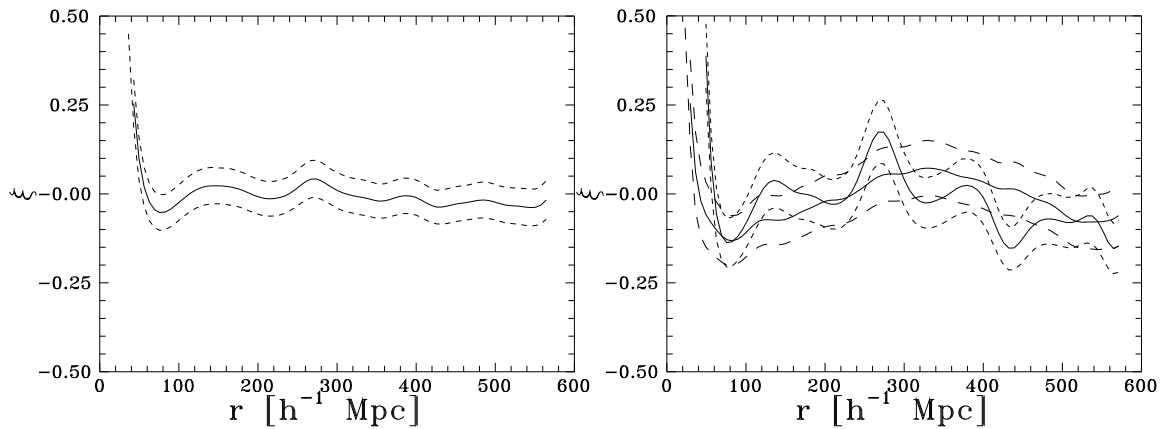}
\caption{
The correlation function of clusters of galaxies with measured
redshifts. The left panel is for the sample of all clusters
(ACO.R.H1). In the right panel data on high- and low-density
populations are given separately. Solid lines show the correlation
function for the samples ACO.R.H4 and ACO.R.L4; the error corridor for
high- and low-density cluster populations is marked with short and
long-dashed lines, respectively. The overall curved shape of the
correlation function is due to cosmic variance (compare with Figure~7
of Paper III).}
\label{fig2}
\end{figure*}

\begin{figure*}
\vskip -17mm
\epsfysize=7cm
\epsfbox [-10 30 300 260]{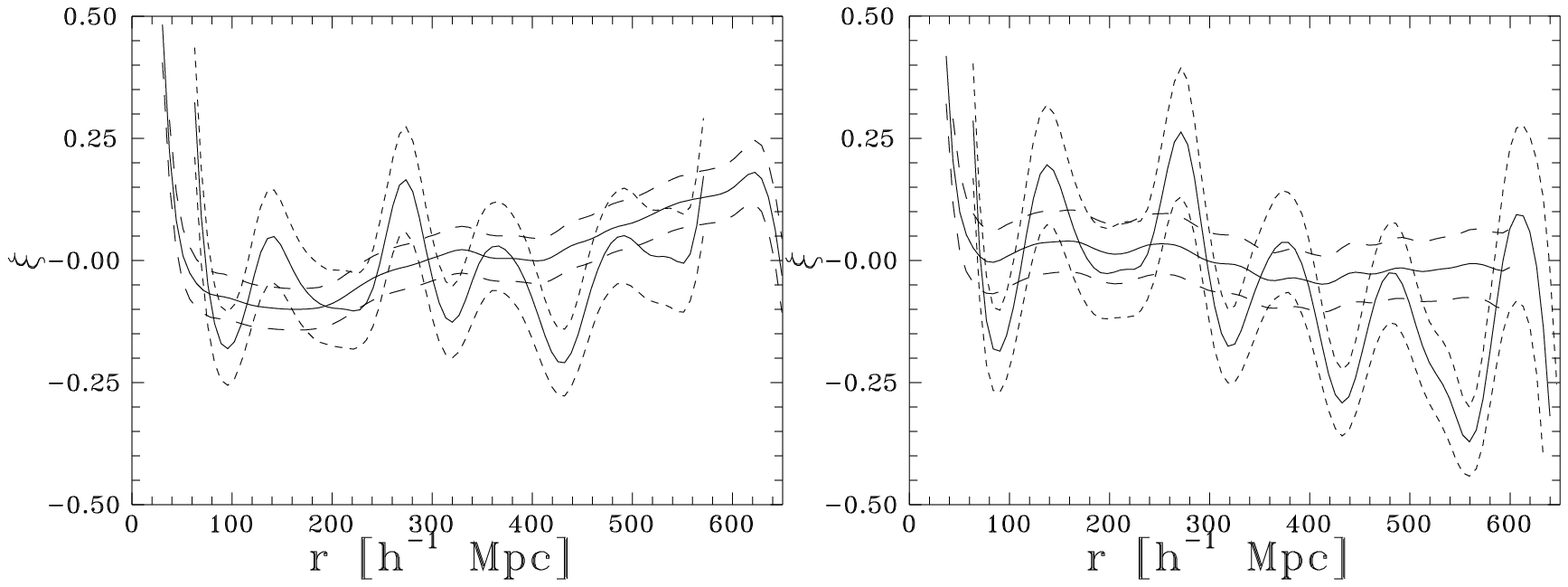}
\caption{
The correlation function for all clusters 
(samples ACO.A.H8 and ACO.A.L8 in the left panel), and for
clusters with measured redshifts (samples
ACO.R.H8 and ACO.R.L8 in the right panel).  
Solid, dashed, and dotted lines have the same meaning
as in Figure~2.}
\label{fig3}
\end{figure*}

To determine the distance dependence of the selection function the
spatial density of clusters of galaxies was calculated in concentric
spherical shells of thickness 20~\Mpc, for each hemisphere separately.
Fluctuations are rather large, thus for this sample of clusters the
mean regression was derived for both hemispheres. The spatial density
can be represented by a linear law: $D(r)=d_0-d_1(r/r_1)$, where $d_0$
and $d_1$ are constants, and $r_1$ is the outer radius of the
sample. Values of the selection function parameters  $d_0$, and
$d_1$, found for various subsamples of clusters, are given in
Table~1. 

A similar analysis of the selection function was made for the sample
of all 1304 clusters. Here, too, the sample was divided into high- and
low-density populations using the same threshold $N_{cl}=8$.  Table~1
shows that parameters of the distance dependence in the Northern and
Southern hemisphere (denoted with subscripts $N$ and $S$,
respectively) are identical in most cases. Only the cluster sample of
all clusters in low-density regions is large enough to determine
parameters of the distance dependence separately for both hemispheres.
Here $d_{0N}$ is smaller than $d_{0S}$, which reflects the fact that
the number-density of the Northern cluster sample is lower than that
of the Southern one.  Parameters for the selection effect in Galactic
latitude are similar for the sample of all clusters and that of
clusters with measured redshifts.

\section{  THE CLUSTER CORRELATION FUNCTION}

\subsection{  Deep cluster samples}

In this Section we discuss the correlation function of Abell--ACO clusters of
galaxies in various environments.  As noted above, clusters in high-density
environment (rich superclusters) form a fairly regular three-dimensional
network, whereas clusters in low-density environment (isolated clusters and
clusters in poor and medium rich superclusters, or simply poor superclusters)
are located in their vicinity more irregularly (Paper~I). To determine which
limiting richness $N_{cl}$ divides clusters naturally into high- and
low-density environment, we calculated the correlation functions for both
populations using limiting richnesses between $N_{cl}=1$ and $N_{cl}=8$.  For
$N_{cl}=1$ per definition there are no clusters in the low-density population
(since low-density population consists of clusters in superclusters of
multiplicity less than $N_{cl}$). Results for $N_{cl}=1, ~4$, and 8 are shown
in Figures~2 and 3 for clusters with measured redshifts.

These Figures show that the correlation function of clusters in
rich superclusters has a number of quasi-regularly spaced
secondary maxima and minima (in addition to the main maximum at small
separation).  This phenomenon is the main finding of the present paper
and we shall refer to it as the {\it oscillation} of the correlation
function.

In contrast to the correlation function of clusters in rich
superclusters the correlation function of clusters in poor
superclusters approaches zero smoothly after the initial
maximum. The nearest neighbour test and void analysis show (Paper I) that
clusters in poor superclusters are located more irregularly in void walls
between rich superclusters and thus secondary peaks of the correlation
function due to individual poor superclusters cancel each other out.

Parameters of the oscillations of the correlation function for clusters in
rich superclusters are given in Table~2: $N$ is the number of clusters in the
sample; $r_{min}$ is the location of the first secondary minimum of the
correlation function; $r_{max}$ is the location of the first secondary
maximum; $A_{max}$ is the amplitude, which is defined as half of the
difference of the value of the correlation function between the first
secondary maximum and minimum; $\sigma_{\xi}$ is the mean $1\sigma$ error of
the correlation function, which determines the width of the error corridor;
$\Delta_{21}$ and $\Delta_{32}$ are distances between secondary maxima
indicated by respective indices; and $\Delta_{mean}$ is the mean separation of
the secondary maxima, and of the secondary minima.  Positions of the maxima
and minima and differences between them are given in \Mpc.  The mean error was
calculated from Eq. (16) of Paper III. Essentially the error is determined by
the cosmic variance (i.e. the variation of the correlation function in
different volumes of space):
$$
\sigma_{\xi c}={b \over \sqrt{N}}, \eqno(1)
$$
where $b$ is a parameter introduced in Paper III to describe the
dependence of the error on the character of the large-scale
distribution of clusters of galaxies. It must be determined from mock
samples. We have done this (for details see  Paper III) and found that
$b\approx 1.5$, see also the discussion in Sect. 4. As we see from the above
equation, the width of the error corridor for the cosmic variance is
constant. 

We see from Table~2 that the amplitude of oscillations increases with the
increase of the minimum supercluster richness $N_{cl}$.  This leads us to the
conclusions that, for low values of $N_{cl}$, we actually have a mixture of
populations in the high-density population, and that the proper division of
populations occurs at the highest minimum richness, $N_{cl}=8$. To check this
result we have calculated the correlation function separately for clusters
located in superclusters of medium richness, from $N_{cl}=4$ to $N_{cl}=7$.
The correlation function of this subpopulation shows only marginal signs of
oscillations. Thus we can accept $N_{cl}=8$ as the limiting richness to select
the regularly distributed population of clusters in rich superclusters.  This
analysis confirms results found in Papers I and III: a smooth distribution in
void walls leads to a non-oscillating correlation function in the case of
clusters in poor superclusters; oscillations occur only in the case if rich
superclusters are located in a quasi-regular rectangular lattice.

In Table~2 we give parameters of the oscillating correlation function
for the cluster population with measured redshifts.  The sample of all
clusters was also divided into high- and low-density populations, and
parameters of the correlation function were determined. Results for
samples with measured redshifts and for all clusters are given in
Figure~3. In this case we see that, on large scales, clusters in
rich superclusters have an oscillating correlation function and
clusters in poor superclusters have a zero correlation.  Parameters
of the oscillations of clusters in rich superclusters have values
very close to values for the sample of clusters with measured
redshifts; only the amplitude of oscillations is smaller by a factor
of about 1.5.  A smaller amplitude for the sample of all clusters is
likely due to the larger observational errors in the photometric
redshifts, which smooth out features slightly in the correlation
function.

\begin{figure*}
\epsfxsize=280pt
\vskip -17mm
\epsfysize=7cm
\epsfbox [-10 30 300 260]{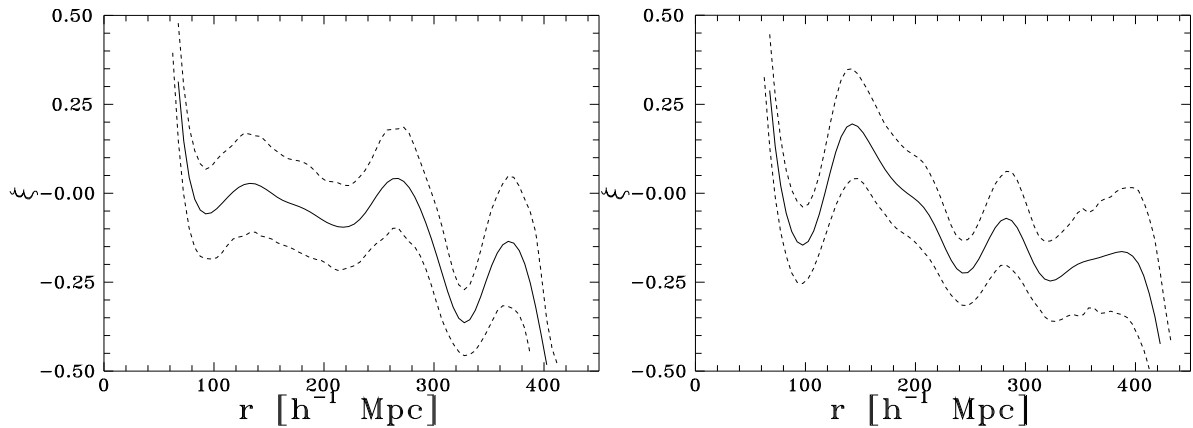}
\caption{
The correlation function calculated separately for the Northern (left panel)
and Southern (right panel) Galactic hemispheres for all clusters 
located in rich superclusters. Error corridors are also given.}
\label{fig4}
\end{figure*}

Now we compare the error in the correlation function for subsamples
with various limiting richness $N_{cl}$. We see that the amplitude of
oscillations for the sample ACO.R.H8 is approximately three times
larger than the error; i.e., we are able to establish the presence of
oscillations at a $3\sigma$ level. For the sample of clusters of all
richness classes taken together (ACO.R.H1) the error is approximately
equal to the amplitude of oscillations.  This shows that the division
of clusters into high- and low-density populations is crucial {\it to
demonstrate the presence of oscillations}. (We note, however, that the
power spectrum of the cluster population in rich superclusters is
almost identical in shape to the spectrum of the whole cluster
population.)

\subsection{ Cluster samples in the Northern and Southern hemispheres }

Now we determine the cluster correlation function separately for the
Northern and Southern Galactic hemispheres. To increase the number of
clusters we use the sample of all clusters, and divide this sample
again into rich and poor superclusters using the limiting richness
$N_{cl}=8$.  Figure~4 shows the correlation function of clusters
located in rich superclusters separately for both Galactic
hemispheres. We see that there are some differences between the
correlation functions. 

The oscillatory behaviour is very clear in both cases, and the period of
oscillations is identical (see Table~2). The basic difference lies in the
amplitude, which is smaller for the Northern hemisphere. This suggests that
the supercluster-void network is less regular in the Northern hemisphere. It
is interesting to note that Landy \etal (1996) have determined the power
spectrum of galaxies in the deep Las Campanas Redshift Survey separately for
the Northern and Southern Galactic hemispheres.  The Southern samples have a
strong peak at a wavelength $\approx 100$ \Mpc, whereas in Northern samples
this feature is much weaker. The similarity of these independent measures of
the regularity of the structure suggests, first of all, that both methods (the
correlation and spectral analyses) work and that they measure the large-scale
regularity of the structure. Secondly, these results indicate that there are
small-but-definite differences in the large-scale distribution of high-density
regions in the nearby Universe.  In other words, Northern and Southern
samples, taken separately, do not form fair samples of the Universe.

\begin{table*}
\begin{minipage}{120mm}
\caption{ Parameters of the correlation function for 
various cluster samples}
\halign to \hsize {# \hfil &\quad\hfil #&\quad \hfil#\hfil&\quad\hfil
#\hfil&\quad \hfil#\hfil&\quad \hfil#\hfil&\quad \hfil#\hfil&\quad\hfil
#\hfil&\quad \hfil#\hfil\cr
\noalign {\smallskip}
Sample&$N$&$r_{min}$&$r_{max}$&$A_{max}$&$\sigma_{\xi}$&
$\Delta_{21}$&$\Delta_{32}$&$\Delta_{mean}$\cr
\noalign {\medskip}
ACO.R.H1    & 869&78&144&0.056&0.051&126&116&122\cr
ACO.R.H2    & 624&79&131&0.056&0.060&137&117&122\cr
ACO.R.H4    & 433&78&136&0.134&0.072&134&108&123\cr
ACO.R.H6    & 331&83&140&0.200&0.082&132&108&120\cr
ACO.R.H8    & 261&88&138&0.279&0.093&133&104&116\cr
\noalign{\smallskip}
ACO.A.H8$_N$& 152&94&133&0.069&0.130&132&103&118\cr
ACO.A.H8$_S$& 167&97&143&0.275&0.124&140&105&118\cr
}
\end{minipage}
\end{table*}
    
\subsection{ Mean parameters of oscillations}

The grid size of the supercluster-void network can
be determined from data given in Table~2 using relations between the
grid size and parameters given in Paper III.  All scaling parameters depend
on the period $P$ which is equal to the grid size of the supercluster-void
network (see Section 4.4 of Paper III).  The most accurate value of
the period comes from the relation $P=\Delta_{mean}/1.01$; here
$\Delta_{mean}$ is the mean separation between maxima and between
minima. We get
$$
P=115 \pm 15~ h^{-1} {\rm Mpc}. \eqno(2)
$$
The variance of the mean period is given mainly by the error of
positions of the last maximum and minimum.  The error in the location
of the outermost extrema is 25~\Mpc\ which contributes an error of
5~\Mpc\ in $P$.  The actual error is larger as we must take into
account also possible cosmic scatter of the grid size in different
volumes. Comparison of different subsamples yields the error estimate
given in (2).  We note that the value of the period of oscillations is
very close to the mean separation between  rich superclusters
located on opposite sides of voids. The latter separation was found to
be 120~\Mpc\ in Paper I.

The amplitude of oscillations is given by the amplitude of the first
secondary maximum for clusters with measured redshifts located in
rich superclusters:
$$
A=0.28 \pm 0.05. \eqno(3)
$$
The error of the amplitude is estimated on the basis of the scatter
of estimates of the amplitude for different subsamples and of the Poisson
error of data.

\subsection{ The parameters of the correlation function}

Here we determine the
numerical relations between various parameters of the correlation
function.  As demonstrated in Paper III, the separation of the first
secondary maximum of the correlation function from zero is always
larger than the period of oscillations, and the difference between the
second and first secondary maximum is always larger than the
difference between the third and second secondary maximum.

Using the observed correlation function parameters in Table~2 we found
the following relations: $f_1={r_{max}/P}=1.20$;
$f_{21}={\Delta_{21}/P}=1.16$; and $f_{32}={\Delta_{32}/P}=0.84$;
where $\Delta_{21}$ and $\Delta_{32}$ are mean separations of
respective maxima of the correlation function.  A comparison of
numerical values for these  parameters with respective
values found for model samples in Paper III shows rather close
agreement.  This agreement is an additional argument indicating the
reality of our results.

\subsection{ The correlation length}

In this paper the major emphasis is on the study of the correlation
function of clusters of galaxies on large scales. Our data contain
information also on the correlation function on small scales, and in
this Section we discuss our results for the determination of the
correlation length. This parameter is defined as the value of the
separation $r=r_0$ at which the correlation function $\xi(r_0)=1$.
This parameter depends critically on the 
characteristic size of superclusters.

We determined the correlation length using non-smoothed correlation
functions since smoothing increases it. As for
other parameters, the correlation length was found separately for
cluster samples in rich and poor superclusters. Results are
interesting: for clusters in rich superclusters 
the correlation length is
$$
r_0=46 \pm 5~ h^{-1} {\rm Mpc}, \eqno(4)
$$
and for clusters in poor superclusters
$$
r_0=17 \pm 3~ h^{-1} {\rm Mpc}. \eqno(5)
$$
The errors are estimated on the basis of the scatter from  
samples for various minimum multiplicity. 
Differences in the correlation function at small scales are seen
also in Figures~2 and 3, although the smoothing makes the correlation 
length appear larger. 

These differences are expected when we take into account the geometric
meaning of the correlation length -- it is close to the mean minor
diameter of systems of clusters. Poor superclusters are small, but
rich ones have much larger diameters (Jaaniste \etal 1997).  Similar
differences are found also for clusters in rich and poor superclusters
in models (Paper III).  These calculations show that there exists no
unique correlation length for clusters; it is in fact a function of
cluster environment (the size of superclusters).

\section{ TESTING THE REALITY OF OSCILLATIONS}

The presence of oscillations in the cluster correlation function was
first established by one of us (VS) in December 1994 and presented in a
preprint by Saar \etal (1995). Since then we have discussed this result
at several conferences and seminars. During these discussions a number of
questions were raised: Perhaps the local minima and maxima of the
correlation function are just a random noise or due to selection
effects, supercluster definition, smoothing, or some other disturbing
effect? And if oscillations are real, can they be reproduced in the
framework of conventional CDM cosmogony with Gaussian initial
fluctuations, or do they demand a radical change of our paradigms on
the formation of structure in the Universe? To answer these questions
we have performed a number of tests. In this Section we discuss
the reality of oscillations.

\subsection{ Errors in the correlation function}

The most serious question is related to errors in the correlation
function.  Often the errors in the correlation function are calculated
from Poisson statistics. Mo, Jing, \& B\"orner (1992) have shown that
the cosmic variance is much larger than the Poisson noise, and our
results have confirmed this.  Einasto \& Gramann (1993) determined the
error corridor by a bootstrap procedure. This method is also not very
accurate since it cannot handle real variance of samples in different
volumes of space. The only way to get an idea of the possible effect
of this cosmic variance is to study various models of the cluster
distribution.

Results of this study are presented in detail in Paper III. It is shown
that the error corridor of the correlation function due to cosmic 
variance depends on the size of the sample (the number of particles
$N$) and the nature of the distribution of particles, and can be
parameterised by Eq. (1) presented above. The parameter $b$ of
this equation has a value  about $b\approx1.5$ in  models which have a
large-scale distribution of clusters similar to the observed
distribution. In our calculations we have used this value of the error
parameter. The amplitude of oscillations of the correlation function
for the subsample of clusters in rich superclusters ACO.R.H8 is
about 3 times larger than the error; thus cosmic errors do not play an 
important role. If we use the sample of all clusters with redshifts 
 (ACO.R.H1) then the amplitude of the correlation function is
approximately equal to the cosmic variance (cf.  Figure~2). Thus it is
essential to divide the cluster sample into two populations with
different properties of the spatial distribution to establish the
oscillatory behaviour of the cluster correlation function.

\begin{figure*}
\vskip -17mm
\epsfysize=7cm
\epsfbox [-10 30 300 260]{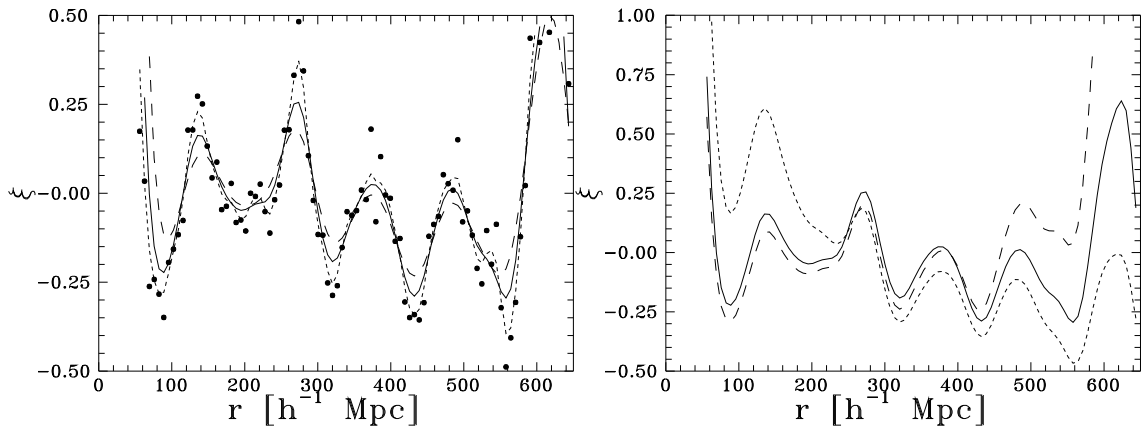}
\caption{
The influence of procedural artefacts on the correlation function. 
In the left panel for the sample with redshifts in rich superclusters dots
show the unsmoothed correlation function; dotted, solid, and dashed
lines show the correlation function smoothed with Gaussian dispersion
6.5, 13, and 20 \Mpc, respectively. In the right panel the influence of
the selection function in the correlation function is given. The dotted
line is for the selection function with parameters $s_0=0.14$,
$d_0=1$, $d_1=0.50$; the solid line is for selection function
$s_0=0.38$, $d_0=1$, $d_1=0.80$ (correct values); and the dashed line
for selection function with $s_0=0.38$, $d_0=1$, $d_1=0.90$.}
\label{fig5}
\end{figure*}

\subsection{ Sample shape}

Since the sample volume has the form of a double cone and is restricted to a
limiting distance, we will now check whether the curious shape of the sample
can artificially generate oscillations in the correlation function.

The strongest evidence against such an effect comes from the
comparison of samples in rich and poor superclusters (cf Figures~2,
3).  Both samples occupy identical double-cone shaped volume. The only
difference lies in the spatial distribution of clusters {\it within}
the double conical volume.  It is very difficult to assume a selective
influence of the sample volume shape so that in case of clusters in
rich superclusters the shape generates oscillations in the correlation
function and in case of clusters in poor superclusters it produces a
smooth correlation function near zero. The difference must be
intrinsic.

To check this problem we studied in Paper III the influence of the sample
shape on the correlation function.  Results show that the double conical
sample has about a factor of 4 times fewer particles than the whole cubical
sample, and thus cosmic variance is larger, but the value of the error
parameter $b$ is almost the same as for the whole cubical sample. In the cases
in which structural elements (clusters in high-density regions) led to an
oscillatory behaviour of the correlation function, these were present in
sufficient quantity also when restricting the sample volume to a double-cone.
If the size of the conical sample is very small, then characteristic elements
which determine the oscillating properties of the correlation function are not
present in sufficient quantities and the correlation function becomes
irregular.

\subsection{ Supercluster selection}

The supercluster catalogue used in this study was compiled in Paper I using a
neighbourhood radius 24 \Mpc.  Is this radius crucial for the oscillatory
behaviour of the correlation function?

The dependence of the supercluster catalogue on the neighbourhood radius was
investigated by Einasto \etal (1994). For neighbourhood radii $\ge 32$~\Mpc\ 
almost all clusters join to form one huge percolating system. Thus it is clear
that a meaningful neighbourhood radius must be smaller than this value. If the
radius is very small then we select as superclusters only the highest density
peaks of the distribution of clusters, and the number of clusters in
superclusters becomes too small for the determination of the correlation
function.  To determine the influence of this parameter we compiled
superclusters using a series of values of the neighbourhood radius: 12, 16,
and 20 \Mpc. For all cases the correlation function for clusters was
calculated. The results indicate that with decreasing neighbourhood radius the
amplitude of oscillations of the correlation function increases since only
very compact superclusters will be selected. However, positions of the maxima
are practically the same as for the adopted neighbourhood radius (24~\Mpc).
This test shows that the oscillating behaviour and parameters of oscillations
are quite stable and do not depend on the choice of the neighbourhood radius.

\subsection{ Smoothing scale}

To investigate the influence of the smoothing length on our results we
calculated the correlation function for one sample with various values
of the dispersion $\sigma_{s}$. Results are shown in Figure~5. This
calculation shows that there is no principal difference between
results for different smoothing lengths. Main parameters of the
correlation function (the period and positions of the maxima and
minima) change only within a few per cent. The largest change is in the
amplitude of oscillations, which decreases considerably with the
increase of the smoothing length. To avoid the influence of the
smoothing  we determined the amplitude from non-smoothed data.
In all Figures we have used smoothing length, $\sigma_{s}=13 - 15$ \Mpc.
This almost completely removes the Poisson noise, and is sufficient
to investigate details of the correlation function above a scale of 30
\Mpc.

\subsection{ Selection function}

One frequently asked question is the influence of the selection function. If
the feature investigated is of the same scale as the depth of the sample then
small errors of the selection function can seriously influence the results. To
investigate the influence of the selection function in our case we calculated
the correlation function of one sample for a number of different selection
function parameters used in the calculation of comparison Poisson samples.
Results are presented in Figure~5. In all cases the same procedure was applied
to calculate the selection function (discussed in Section 2 above).  Only the
parameters of the selection function were changed. As test sample we chose
clusters in rich superclusters (ACO.R.H8). In this case the number-density of
clusters decreases very rapidly with increasing distance from the Galactic
pole (cf. Figure~1). If we ignore this rapid decrease and adopt a standard
value for the selection parameter (as for all clusters), $s_0=0.14$, then the
overall mean slope of the correlation function changes. If we change the
parameter which determines the decrease of the number-density of the sample
with distance and adopt too low a value for the number-density on the far side
of the sample ($d_1=0.9$ instead of the correct value $d_1=0.8$), then the
whole correlation function on large scales increases. Both changes of
selection function parameters have, however, little effect on the main
parameters of the correlation function: none of the parameters quoted in Table
2 change by more than a few per cent. Thus we can say that small errors of the
selection function do not influence our main results. This insensitivity is
due to the fact that the size of our sample is much larger than the scale of
interest.

\begin{figure}
\vskip -10mm
\epsfysize=7cm
\epsfbox [45 40 300 300]{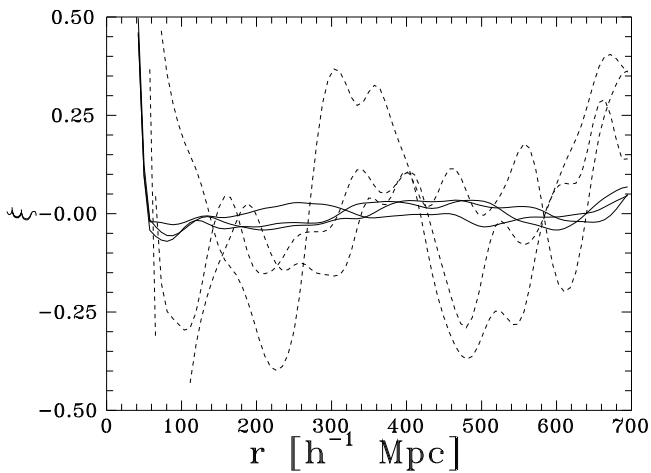}
\caption{
The correlation functions of clusters for CDM models.  Solid lines
are for clusters in double conical subsamples located in poor
superclusters (with less than 8 members) and dotted lines are for clusters
in double conical volumes in rich superclusters (with at least 8
members).}
\label{fig6}
\end{figure}

\section{COMPARISON WITH MODELS}

In this Section we compare our empirical correlation function of
clusters of galaxies with correlation functions calculated for several
models. We use  CDM-models of structure evolution, models with a double
power-law spectrum, as well as geometrical models with randomly
and regularly located superclusters. Our main questions are: Can the
observed correlation function of clusters of galaxies be reproduced by
conventional models of structure evolution? If not, what changes in models
are needed to reproduce the observed function?

\subsection{ Comparison with CDM-models}

We have calculated several $N$-body models of structure evolution. One model
is based on the standard CDM-scenario of structure formation.  It has the
structure parameter $\Gamma= \Omega h=0.5$, with the Hubble parameter $h=0.5$,
and the density parameter $\Omega=1$.  The second model was calculated with a
double power-law perturbation spectrum, with spectral index $n=1$ on large
scales (wavenumber $k<k_0$), index $n=-1.5$ on small scales (wavenumber
$k>k_0$), and transition at wavelength $\lambda_0=2\pi/k_0=115$ \Mpc.  Models
were calculated using a particle-mesh code with $128^3$ particles and $256^3$
cells in a cube of size $L=700$ \Mpc.  Clusters of galaxies were searched with
a method similar to the ``friends-of-friends'' algorithm. The mass of clusters
is determined from the number of particles in volumes of enhanced density.
The lower limit of the mass of clusters was chosen so that the total number of
clusters in the sample was in agreement with the mean spatial density of
Abell--ACO clusters.

\begin{figure*}
\epsfysize=7cm
\epsfbox [15 375 300 600]{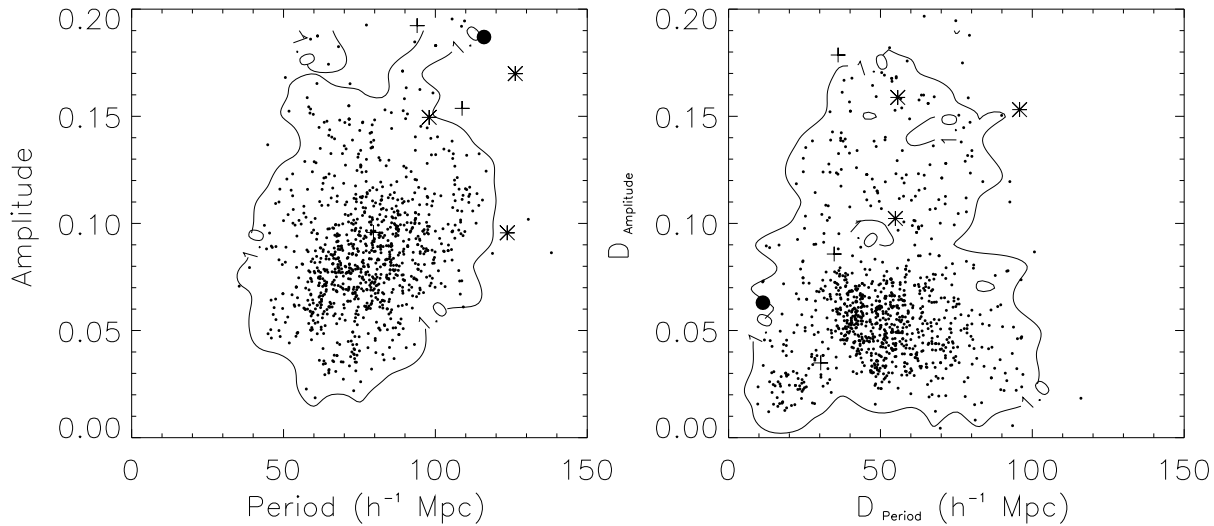}
\caption{
Parameters of oscillations of the correlation function: period,
amplitude (left panel), and their scatters, $D_{period}$, $D_{ampl}$
(right panel). The large filled circle shows the observed values for
clusters in rich superclusters (sample ACO.R.H8); dots are
respective values for 1000 realizations of the random supercluster
model, crosses for the standard CDM model, and stars for a low-density
CDM model with cosmological constant (see Paper III for details).
Contours indicate the probability level for random superclusters
outside of which 1 \% of periods and amplitudes are found. To
calculate parameters of oscillations for this figure we used smoothed
correlation functions. In this case the amplitude of oscillations from
observations is $A=0.186$ (the value given in Table~2 corresponds to
the amplitude of the unsmoothed correlation function).}
\label{fig7}
\end{figure*}

We calculated the correlation function of model clusters for the whole
box using all clusters and also for double conical subsamples of
clusters in rich and poor superclusters. We applied a supercluster search
algorithm identical to the one used for the search of real
superclusters with neighbourhood radius 24 \Mpc. In each
of our simulations we constructed three double conical volumes (cone
axes directed along the three axes) and searched clusters in these
volumes. Clusters were divided into two populations -- one in rich
superclusters and the other in poor ones, with limiting richness
$N_{cl}=8$ as in the real case.  Correlation functions found for the
CDM model are plotted in Figure~6.

There are no regular oscillations in the correlation function in rich
superclusters in either the whole cubical sample or in the double conical
volumes.  The correlation functions of simulated clusters in the double
conical volumes and located in rich superclusters have several peaks and
valleys on large scales, but the location and amplitude of these peaks is
random (for details see next subsection). Model clusters in poor
superclusters have a smooth correlation function close to zero at large
scales.

This result is expected as the power spectrum of CDM-models is a smooth
function of wavenumber, with a continuous change in the slope of the
spectrum. For such spectra  oscillations of the correlation function
are not expected since oscillations occur only in the case when the
spectrum has a peak and the slope near the peak changes suddenly
(Frisch \etal 1995, Paper III).

This does not exclude the possibility that, in some realizations of a
model with a CDM-type perturbation spectrum, peaks and valleys in the
correlation function of clusters in rich superclusters are located
more regularly. This occurs when the perturbation spectrum accidently
has an extra peak near its maximum.  In the next subsection we study 
more closely the possibility of how frequently such a peak can occur.

\subsection{ Comparison with random supercluster samples}

To investigate the possible generation of regular oscillations in the
correlation function for double conical volumes of clusters in
rich superclusters we must generate a large number of
realizations of  models.  The distribution of clusters in models is
determined essentially by  medium scale perturbations which are still
in the linear stage of evolution. Thus it is not necessary to use
conventional $N$-body calculations of structure evolution. Borgani
\etal (1995) have used the Zeldovich approximation for a similar task.
In this paper we shall apply an even simpler procedure to investigate the
regularity of the large-scale distribution of clusters.

In the present problem it is not essential to use exactly the CDM
spectrum.  What is important is to apply a broad band spectrum with a
smooth transition between regions on large and short wavelengths.  As
demonstrated in Paper III, the power spectrum of the random
supercluster model is rather similar to the power spectrum of
CDM-models, in particular in the medium wavelength region of interest
for the present study. Correlation functions of these models are also
very similar.  We make use of this similarity and generate a large
number of realizations for the random supercluster model to see how
frequently such a model can reproduce properties of the real
correlation function.

In this model (for details see Paper III) superclusters are located
randomly in space. They contain clusters of galaxies in a number which
is in agreement with the observed multiplicity function of
superclusters. To imitate the observations we choose a double conical
sub-volume from the whole cubical sample and select clusters which
belong to rich superclusters with at least 8 member clusters. The full
side length of the cube is taken to be $L=700$ \Mpc. The number of
superclusters in models is taken to be approximately equal to 650; in
this case the number of clusters in rich superclusters of double
conical subsamples is about 300 as in the observed cluster sample in
rich superclusters.  Our calculations show that the correlation
function of this model also has maxima and minima, but they are
located randomly, similar to the cluster correlation function of CDM
models. We can characterise oscillations and their regularity by the
following parameters: the mean period of oscillations, its rms
scatter, the mean amplitude of oscillations, and its rms scatter.

\begin{figure}
\epsfysize=7cm
\vskip -10mm
\epsfbox [50 45 300 300]{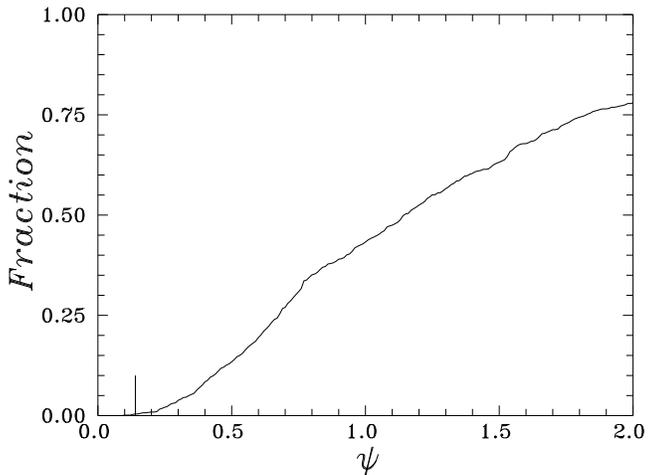}
\caption{
The integrated frequency distribution of the correlation function 
variance parameter $\psi$.
The observed value of $\psi$ is noted by a vertical bar.}
\label{fig8}
\end{figure}

Results of our calculations for 1000 realizations of the random
supercluster model are shown in Figure~7, separately for the amplitude
vs. period and for the scatter of the amplitude vs. the scatter of the
period.  If a point lies outside the 1~\% contour, it has a
probability of occurrence of $<$1~\%.  We see that for both variable
parameter pairs the observational point lies just outside the 1~\%
contour. In other words, the probability that our observed sample is
taken from the same model is approximately 1~\% for both variable
pairs.

We applied a further test using the fine details of the correlation
function. As noted above, the position of the first secondary maximum
of the correlation function, as well as mean differences between the second
and first, and between the third and second maxima, are in certain 
fixed relations with the period of oscillation. We can define a
correlation function variance parameter as follows:
$$
\psi^2=(f_0-f_{00})^2 + (f_1-f_{10})^2 + (f_2-f_{20})^2, \eqno(6)
$$
where $f_0$, $f_1$, and $f_2$ are values of parameters defined by Eqs
(12) -- (14) of Paper III and found for the test model; $f_{00}$,
$f_{10}$, and $f_{20}$ are respective values calculated for the
geometric model with regular structure. As demonstrated in Paper III,
these parameters are rather stable and depend only little on models
with different details of the structure. Essential is the presence of
a regular network of superclusters and voids. Thus we have calculated the
correlation function variance parameter $\psi$ for all our 1000 test
models (see Figure~8).

This calculation shows that the mean value of the parameter is
$\psi=1.4$. The distribution is very asymmetric with a long tail
towards large $\psi$ values. The lowest value for these 1000
realizations is 0.1.  The observed value is $\psi\approx 0.14$.  We
see that the probability that the observed case is taken randomly from
the family of random supercluster model is also about 1~\%.  All our
variables used in these tests are independent of each other, thus the
probability to get all five parameters fitted once by the random
supercluster model simultaneously is much smaller than 1 \%.

Even if using the random supercluster model is a fast but not ideal procedure
for calculating these probabilities the main result would be hardly changed by
more ingenious simulations: the probability is very small.  Thus we conclude
that within standard cosmological models it is difficult to generate the
observed correlation function.

\begin{figure}
\epsfysize=7cm
\vskip -10mm
\epsfbox [50 45 300 300]{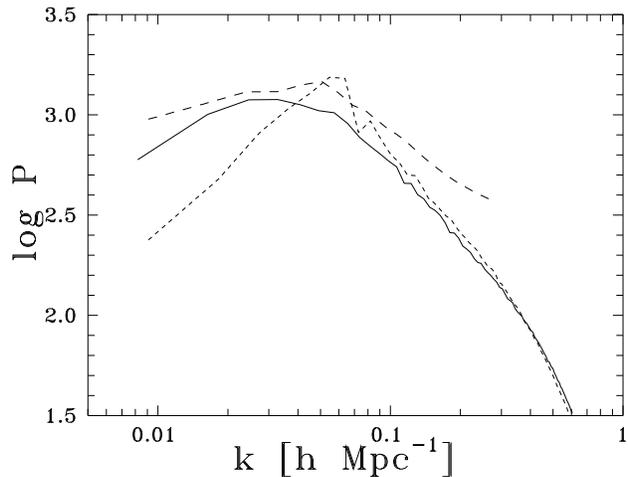}
\caption{
Power spectra  for the CDM model, the double power-law model, and the 
mixed geometrical model, plotted with  solid, dotted, and 
dashed lines, respectively. }
\label{fig9}
\end{figure}

\section{ POWER SPECTRUM}

Which perturbation spectrum can produce the observed correlation function of
clusters in rich superclusters?  Analytic calculations made in Paper III show
that the correlation function has an oscillatory behaviour only if the power
spectrum has a peak at the corresponding wavenumber. In this paper it was also
demonstrated that the sharpness and height of the peak in the spectrum
determines the character of oscillations of the correlation function.

Here we estimate the possible shape of the spectrum on scales of
interest using comparison with models with known spectra.
We shall compare the spectra and correlation functions of three
models: the standard CDM model, the double power-law model, and a
mixed geometrical model consisting of two populations, one with
superclusters located randomly along regularly spaced rods and the
other of irregularly spaced superclusters (see Paper III for
details). Power spectra of these three models are shown in Figure~9.

We see that the double-power law model and the mixed model have rather
similar spectra near the maximum. Both models have also similar
correlation functions with weak oscillations (see Figure~4 of Paper
III).  The oscillations are more regular in the geometrical model, as
expected. However, the differences between models are not large.  The
maximal deviation of the spectrum near the maximum from the
corresponding CDM-type spectrum is  $\sim 0.2$ dex,
i.e. about a factor of 1.25 in amplitude. 

These models show that already a modest deviation from the
standard CDM spectrum  produces an oscillating correlation
function for clusters in rich superclusters.

The actual power spectrum of our cluster sample has a peak of even
higher amplitude (see E97).

\section{  CONCLUSIONS}

We have determined the correlation function for clusters of galaxies
separately for all clusters and for clusters located in rich and in
poor superclusters. The correlation function of clusters in rich
superclusters that form the skeleton of the supercluster-void network
has an oscillatory behaviour with a period of $115
\pm 15$ \Mpc. Within an interval of $\sim650$ \Mpc\ over which 
observational data are available, five secondary maxima and minima of
the correlation function are seen.  The amplitude of oscillations is
larger for clusters located in very rich superclusters.

The scale of the supercluster-void network found here on the basis of the
cluster correlation function is rather close to the scale found using other
methods, such as void diameter analysis, pencil-beam studies, or absorbers in
the line-of-sight to QSOs (Quashnock et al. 1996), although the latter applies
to higher redshifts.

The reality of oscillations of the cluster correlation function
is supported by the following arguments.

(1) The error corridor of the correlation function determined for clusters
in rich superclusters is much smaller
than the amplitude of oscillations.
(2) Oscillations are seen in cluster samples located
in both Galactic hemispheres.  
(3) Similar oscillations with lower amplitude are observed in the
Las Campanas Redshift Survey of galaxies by Tucker \etal (1995, 1997).
(4) In all samples the shape of the oscillating
correlation function  follows almost exactly the expected shape
 for a quasi-regular network of superclusters and voids.
(5) The double conical shape of the volume sampled by 
clusters cannot influence the results. 
(6) Parameters of the oscillations practically do not depend
on the smoothing length of the correlation function, nor on the
neighbourhood radius used in supercluster definition, or on errors of
the selection function used to calculate the correlation function.

 The correlation length of clusters of galaxies depends on the cluster
population: for clusters in poor superclusters it is about 17 \Mpc; for
clusters in  rich superclusters it is about 45 \Mpc.

We have compared the observed correlation function with correlation
functions calculated for clusters in CDM-models and for models with
randomly distributed superclusters. These models have a broad-band
power spectrum with a smooth transition between the positive spectral
index at long wavelengths and a negative index at small wavelengths.
In these models  the correlation function of clusters in rich
superclusters located in double conical volumes also has peaks and
valleys, but these peaks and valleys are distributed randomly and have
random amplitudes. The probability that a model with a 
broad-band power spectrum has parameters of oscillations of the
correlation function similar to observed parameters is very low 
($\ll 1$~\%).

Analytical calculations show that oscillations of the correlation
function appear only in case that the power spectrum has a peak at the
wavelength equal to the period of oscillations.  We have compared
spectra and correlation functions of models with various heights of
the peak in the spectrum. These calculations show that it is possible
to generate an oscillating correlation function for clusters in
rich superclusters if the height of the peak is of the order of a
factor of at least 1.25 in amplitude over the conventional smooth
spectrum.

The fact that the amplitude of oscillations near the last maximum is
still rather large suggests that the coherence of positions of
high-density regions extends over very large separations (at least
10~\% of the diameter of the observable Universe).

\noindent {\bf Acknowledgements:} 

This work was supported by Estonian Science Foundation grant 182 and
International Science Foundation grant LLF100. We thank Bernard Jones,
Jerry Ostriker and Jim Peebles for discussions. JE and AS were
supported by the Deutsche Forschungsgemeinschaft in Potsdam; AS was
partially supported by the Russian Foundation for Basic
Research under Grant 96-02-17591.


\begin{thebibliography}{99}

\bibitem{a} Abell, G., 1958, ApJS 3, 211
\bibitem{aco} Abell, G., Corwin, H. \& Olowin, R., 1989, ApJS 70, 1 (ACO)
\bibitem{ats1} Andernach, H., Tago, E., \& Stengler-Larrea, E., 1995, 
Ap. Lett.\& Comm. 31, 27 
\bibitem{ats2} Andernach, H., Tago, E., \& Stengler-Larrea, E., 1997, 
(in preparation)
\bibitem{b} Bahcall, N. A., 1991, ApJ 376, 43
\bibitem{bs} Bahcall, N. A., Soneira, R.M., 1983, ApJ 270, 20
\bibitem{be}  Bond, J.R., \& Efstathiou, G., 1984, ApJ 285,
L45 
\bibitem{bpcm} Borgani, A., Plionis, M., Coles, P., \& Moscardini, L.,
1995, MNRAS 277, 1191
\bibitem{beks} Broadhurst, T.J., Ellis, R.S., Koo, D.C., \& Szalay,
A.S., 1990., Nature 343, 726
\bibitem{dbpo} Dekel, A., Blumenthal, G.R., Primack, J.R., \& Olivier,
S., 1989, ApJ 338, L5
\bibitem{dbps} Dekel, A., Blumenthal, G.R., Primack, J.R., \& Stanhill, D.,
1992, MNRAS 257, 715
\bibitem{dt} Doroshkevich, A. G, Tucker, D. L., Oemler, A., Kirshner,
R. P., Lin, H., Shectman, S. A., Landy, S.D., \& Fong, R., 1996, MNRAS
283, 1281
\bibitem{e97} Einasto, J., Einasto, M., Gottl\"ober, S., M\"uller, V., Saar,
V., Starobinsky, A.A., Tago, E., \& Tucker, D., Andernach, H., 
Frisch, P., 1997a, Nature 385, 139 (E97)
\bibitem{eiii} Einasto, J., Einasto, M., Frisch, P., Gottl\"ober, S.,
M\"uller, V., Saar, V., Starobinsky, A.A., Tago, E., \& Tucker, D., 1997b,
MNRAS (Paper III)
\bibitem{eg} Einasto, J., \& Gramann, M., 1993, ApJ 407, 443
\bibitem{egst} Einasto, J., Gramann, M., Saar, E. and Tago, E. 1993,
MNRAS 260, 705
\bibitem{eetda} Einasto, M., Einasto, J., Tago, E., Dalton, G., \&
Andernach, H., 1994, MNRAS, 269, 301
\bibitem{ei} Einasto, M., Tago, E., Jaaniste, J., Einasto, J., \&
Andernach, H.,  1997c, AA (Paper I, accepted, astro-ph/9610088)
\bibitem{fkl} Fetisova, T. S., Kuznetsov, D. Y., Lipovetskij, V. A.,
Starobinsky, A. A. \& Olowin, R. P., 1993, Pis'ma v Astr. Zh. 19,
508 (engl. transl. in Astron. Lett. 19, 198)
\bibitem{f95} Frisch, P., Einasto, J., Einasto, M., Freudling, W.,
Fricke, K.J., Gramann, M., Saar, V., \& Toomet, O., 1995, AA 296, 611
\bibitem{j97} Jaaniste, J., Tago, E., Einasto, M., Einasto, J., \&
Andernach, H., 1997, (in preparation) 
\bibitem{kp} Kaiser, N., \& Peacock, J. A., 1991, ApJ 379, 482
\bibitem{kkf1} Kopylov, A. I., Kuznetsov D. Y., Fetisova T. S. \&
Shvarzman V. F., 1984, Astr. Tsirk. 1347, 1
\bibitem{kkf2} Kopylov, A. I., Kuznetsov D. Y., Fetisova T. S. \&
Shvarzman V. F., 1988, in Large Scale Structure of the Universe,
eds. J. Audouze, M.--C. Pelletan, A. Szalay (Kluwer), 129
\bibitem{ls96}  Landy, S.D., Shectman, S.A., Lin, H., Kirshner, R.P., Oemler,
A.A., \& Tucker, D., 1996, ApJ 456, L1 
\bibitem{mo1} Mo, H.J., Deng, Z.G., Xia, X.Y., Schiller, P., \&
 B\"orner, G.  1992a, AA 257, 1
\bibitem{mo2} Mo, H.J., Jing, Y.P., \& B\"orner, G., 1992c, ApJ 392, 452 
\bibitem{mo3} Mo, H.J., Xia, X.Y., Deng, Z.G., B\"orner, G., \& Fang,
L.Z., 1992b, AA 256, L23
\bibitem{pw} Peacock, J.A., \& West, M.J., 1992, MNRAS 259, 494
\bibitem{q} Quashnock, J.M., Vanden Berk, D.E. \& York D.G., 1996,
ApJ 472, L69
\bibitem{s95} Saar, V., Tago, E., Einasto, J., Einasto, M. \& Andernach,
H., 1995,  astro-ph/9505053
\bibitem{s} Sutherland, W., 1988, MNRAS 234, 159
\bibitem{tm95} Tucker, D.L., M\"uller, V., Gottl\"ober, S., Oemler, A. Jr., 
Kirshner, R.P., Lin, H., Shectman, S.A., Landy, S.D., \& Schechter, P.L., 1995,
Bull. Amer. Astr. Soc. 27, 1365 
\bibitem{to} Tucker, D.L., Oemler, A. Jr., Kirshner, R.P., Lin, H., 
Shectman, S.A., Landy, S.D., Schechter, P.L., M\"uller, V., 
Gottl\"ober, S., \& Einasto, J., 1997, MNRAS (in press)
\bibitem{zw} Zwicky,F., Wild, P., Herzog, E., Karpowicz M. \& Kowal,
C.T., 1961-68,  Catalogue of Galaxies and of Clusters of
Galaxies, Vols. I -- VI. Pasadena, California Inst. Tech.

\end{thebibliography}
\end{document}